\documentclass[prb,aps,showpacs,twocolumn,amssymb,amsmath,amstext,amsfont,noshowkeys,groupedaddress]{revtex4}
\usepackage{graphicx}
\usepackage{theorem}
\usepackage{dcolumn}
\usepackage{bm}
\usepackage{mathrsfs}

\begin{document}

\title{Scintillation and charge extraction from the tracks of energetic electrons in superfluid helium-4}
\author{W. Guo}
\email[Electronic address: ]{wei.guo@yale.edu}
\author{M. Dufault}
\author{S.B. Cahn}
\author{J.A. Nikkel}
\author{Y. Shin}
\author{D.N. McKinsey}
\affiliation{Physics department, Yale University, New Haven, CT 06520, USA}
\date{\today}

\begin{abstract}
An energetic electron passing through liquid helium causes ionization along its track. The ionized electrons quickly recombine with the resulting positive ions, which leads to the production of prompt scintillation light. By applying appropriate electric fields, some of the ionized electrons can be separated from their parent ions. The fraction of the ionized electrons extracted in a given applied field depends on the separation distance between the electrons and the ions. We report the determination of the mean electron-ion separation distance for charge pairs produced along the tracks of beta particles in superfluid helium at 1.5~K by studying the quenching of the scintillation light under applied electric fields. Knowledge of this mean separation parameter will aid in the design of particle detectors that use superfluid helium as a target material.
\end{abstract}

 \maketitle

\section{Introduction}
Superfluid helium is of great interest as a detector material in particle and nuclear physics research. Impurities can easily be removed from helium with cold traps, and at low temperatures all impurities freeze out on the walls of the container. Consequently liquid helium can be made with extreme purity and negligible radioactive background. Due to its unique quasi-particle excitation spectrum, superfluid $^{4}$He can be used to produce ultracold neutrons through the superthermal effect~\cite{Golub-1987}, which has important applications in determining the neutron beta decay lifetime~\cite{Mampe-1993, Doyle-1994, Huffman} and in the search for a neutron electric-dipole moment (nEDM)~\cite{Golub-1994, Balashov, Henry}. It has also been proposed that superfluid $^{4}$He can be used as a target material for the detection of solar neutrinos.~\cite{Lanou-1987, McKinsey-2000, Ju-2007} Furthermore, helium may be used for the detection of Dark Matter in the form of weakly interactive massive particles (WIMPs). For example, experiments using superfluid $^{3}$He as detection medium are currently under study.~\cite{Mayet} The solar neutrino and beta decay events are electronic recoils, the nEDM signal is the $^{3}$He($n,p$)T reaction and WIMP signals consist of nuclear recoils. The energetic recoil electrons or nuclei move in the liquid and lose their energy principally by ionizing and exciting helium atoms along their tracks. As discussed in Section~II, the ionic recombination and reaction of excited helium atoms with the surrounding atoms results in a bright pulse of scintillation light. With an applied electric field, some of the ionized electrons can be extracted and detected. Both the scintillation signal and the electron signal provide useful information about the recoil particles. The field needed to extract the ionized electrons depends on the separation distance between the ionized electrons and the ions. A determination of the mean electron-ion separation distance for charge pairs produced along the track of an energetic particle will help on the design of such helium-based particle detectors.

In this paper, we discuss the determination of the mean electron-ion separation for charge pairs produced by energetic electrons emitted from a beta source in liquid $^{4}$He. The quenching of the prompt scintillation light under applied fields allows us to obtain information on the mean electron-ion separation distance. In Sec. II we discuss the ionization and scintillation processes involved when an energetic electron is stopped in liquid helium. The amount of prompt scintillation light emitted is proportional to the total number of electron-ion pairs that recombine after ionization. In Sec. III Monte Carlo simulations on charge extraction are described. We shall see that the field dependence of the prompt scintillation light strongly depends on the mean electron-ion separation distance. In Sec. IV we describe the design of the experiment and present the results obtained. By comparing the observed quenching of the prompt scintillation light and the simulations, the mean electron-ion separation is estimated to be $560\pm30$~{\AA}. In Sec. V we discuss the possible methods of detecting the extracted electrons. Application of a two-phase helium-based detector for the detection of WIMPs is briefly mentioned. Finally, there is a summary.

\section{Ionization and scintillation in helium}
When passing through liquid helium, an energetic electron causes ionization and excitation of helium atoms along its path. The average rate at which energy is deposited in liquid helium by an electron of several hundred keV is approximately 50~eV/micron.~\cite{Adams} The electron deposits more energy into the liquid toward the end of its track as it slows down. The average energy to produce an electron-ion pair has been measured for an energetic electron to be about 42.3~eV.~\cite{Jesse} It follows that ionization events are on average 850~nm away from each other. The difference of about 18~eV between the average energy of 42.3~eV to produce an electron-ion pair and the helium ionization energy of 24.6~eV goes into excitation of helium atoms and into kinetic energy of ionized electrons. Sato \emph{et al.}~\cite{Sato} calculate that for every ion produced, 0.45 atoms are promoted to excited states. So on average the kinetic energy of the ionized electrons is roughly (18-0.45$\times$20.6)~eV~$\approx$~9~eV. Some of the ionized electrons, however, may gain a relatively large kinetic energy that is sufficient to ionize further helium atoms through subsequent interactions on their own. These electrons are referred to as $\delta$-electrons, and are responsible for the ※hairy§ appearance of charged particle tracks when they are observed in cloud chambers or in photographic emulsions.~\cite{Mott} The number of $\delta$-electrons $F(E_{\delta})$, per g/cm$^{2}$ path length of the incident electron, with energy higher than $E_{\delta}$ is given approximately by~\cite{Rossi,Tenner-1963}
\begin{equation}\label{delta-ray}
F(E_{\delta})\simeq\frac{2{\pi}NZe^{4}}{m_{e}c^{2}A\beta^{2}}\cdot\frac{1}{E_{\delta}},
\end{equation}
where $m_{e}$ denotes electron mass, $c$ is the speed of light, ${\beta}c$ is the incident particle velocity, and $Z$ and $A$ are the charge and mass numbers of the target material. For relativistic electrons (${\beta}\simeq1$) in liquid helium, we estimate that the total number of $\delta$-electrons with energy greater than 1000~eV, produced per cm track length of the primary electron, is less than ten. Compared to the density of electron-ion pairs (about $1.2\times10^{4}$/cm), $\delta$-electrons contribute only a very small fraction of the ionized electrons.

The excited atoms, electrons, and ions quickly thermalize with the liquid helium. The electron, once thermalized, forms a bubble in the liquid typically within 4 ps.~\cite{Hernandez} The He$^{+}$ ion forms a ``helium snowball" in a few picoseconds.~\cite{Ovchinnikov} Due to the Coulomb attraction, most of the electron-ion pairs undergo geminate recombination in a very short time and lead to the production of He$^{*}_{2}$ excimer molecules
\begin{equation}\label{He-e reaction}
(\texttt{He}^{+}_{3})_{\texttt{snowball}}+(\texttt{e}^{-})_{\texttt{bubble}}~\rightarrow~\texttt{He}^{*}_{2}+\texttt{He}.
\end{equation}
Experiments~\cite{Adams} indicate that roughly 50\% of the excimers that form on recombination are in excited spin-singlet states and 50\% are in spin-triplet states. He$^{*}_{2}$ molecules in highly excited singlet states can rapidly cascade to the first-excited state, He$_{2}(A^{1}\Sigma_{u})$, and from there radiatively decay in less than 10 ns to the ground state,~\cite{McKinsey-thesis} He$_{2}(X^{1}\Sigma_{g})$, emitting ultraviolet photons in a band from 13 to 20~eV and centered at 16~eV. As a consequence, an intense prompt pulse of extreme-ultraviolet (EUV) scintillation light is released following an ionizing radiation event. These photons can pass through bulk helium and be detected since there is no absorption below 20.6 eV.

Of the excited helium atoms 83\% are calculated to be in spin-singlet states and 17\% in triplet states.~\cite{Sato} Excited helium atoms with principal quantum number $n\geq3$ can autoionize by the Hornbeck-Molnar process~\cite{Hornbeck}
\begin{equation}\label{H-M process}
\texttt{He}^{*}+\texttt{He}~\rightarrow~\texttt{He}^{+}_{2}+e^{-},
\end{equation}
since the 2~eV binding energy of He$^{+}_{2}$ is greater than the energy to ionize a He($n\geq3$) atom. Additional electron-ion pairs are thus produced. The decay of the resulting singlet excimers contributes to the prompt scintillation light. A small fraction of the excited atoms are in the $2^{1}$P state and may radiatively decay to the ground state, and producing prompt photons as well. The remaining excited atoms are quickly quenched to their lowest energy singlet and triplet states, He$^{*}(2^{1}$S) and He$^{*}(2^{3}$S), and react with the ground state helium atoms of the liquid, forming vibrationally excited He$_{2}(A^{1}\Sigma_{u})$ and He$_{2}(a^{3}\Sigma_{u})$ molecules~\cite{McKinsey-thesis}
\begin{equation}\label{He-He reaction}
\texttt{He}^{*}+\texttt{He}~\rightarrow~\texttt{He}^{*}_{2}.
\end{equation}
There is an activation energy of about 60~meV for the triplet atoms to react with surrounding helium.~\cite{Koymen} The time scale for this reaction to occur has been found to be about 15~${\mu}$s.~\cite{Keto-1974-1,Keto-1974-2} The time scale of the reaction involving singlet atoms has been studied by McKinsey \emph{et al.}.~\cite{McKinsey-2003} The produced singlet molecules He$_{2}(A^{1}\Sigma_{u})$ radiatively decay quickly, which leads to scattered emissions of the 16~eV photons after the prompt scintillation pulse. The time scale of emitting the afterpulse scintillations has been found to be about 1.7~${\mu}$s.~\cite{McKinsey-2003} The radiative decay of the triplet molecules He$_{2}(a^{3}\Sigma_{u})$ to the singlet ground state He$_{2}(X^{1}\Sigma_{g})$ is forbidden since the transition involves a spin flip. The radiative lifetime of an isolated triplet molecule He$_{2}(a^{3}\Sigma_{u})$ has been measured in liquid helium to be around 13~s.~\cite{McKinsey-1999} The triplet molecules, resulting from both electron-ion recombination and from reaction of excited triplet atoms, diffuse out of the ionization track. They may radiatively decay, react with each other via bimolecular Penning ionization,~\cite{Keto-1974-1} or be quenched at the container walls.

Based on the previous discussion, the prompt ($\sim10$~ns) scintillation light is predominantly produced from the recombination of the electron-ion pairs. If some of the ionized electrons are extracted by an applied electric field, the number of recombined charge pairs decreases, leading to a quenching of the prompt scintillation light. Such field-induced quenching of scintillation produced by weak $\alpha$-particle sources in helium has already been observed with about 5\% quenching of the prompt light~\cite{Ito,Hereford} and about 15\% of the total luminescence~\cite{Roberts} at an electric field of 10~kV/cm around 1~K. The fraction of the ionized electrons that can be extracted in an applied field is an important parameter for the design of particle detectors. The ratio of the prompt scintillation light with and without fields provides us an accurate determination of this fraction.

\section{Theoretical model of charge extraction}
As already mentioned, the average energy of the ionized electrons produced along the track of an energetic electron is about 9~eV. These ionized electrons undergo elastic collisions with helium atoms before their energy becomes low enough to be trapped in bubble states. Due to the low energy-transfer efficiency (about $2m_{e}/M_{\texttt{He}}=2.7\times10^{-4}$ per collision), these electrons make many collisions and undergo a random walk. They may move a significant distance from the locations where they originated. Benderskii \emph{et al.}~\cite{Benderskii} have estimated that the mean range of the ionized electrons, assuming 10~eV initial energy, is roughly 1000~{\AA}. An accurate determination of the electron mean range is, however, complicated. At low energies the electron wavelength is of the same order as the interatomic spacing in liquid helium. One needs to consider electrons interacting simultaneously with multiple helium atoms. D.G. Onn and M. Silver have obtained a mean range on the order of 100~{\AA} for electrons of energy about 1~eV.~\cite{Onn} For secondary electrons with high kinetic energies, the cross-sections for inelastic collisions may become dominant, and the electrons can quickly lose their energy by causing excitations and ionizations in helium. As seen in the simulation by Tenner,~\cite{Tenner-1963} the energy of an electron decreases from about 120~eV to below 20~eV in less than 250~{\AA}. The exact distribution of the electron ranges may be intricate. For simplicity, here we assume a Gaussian distribution of the electron ranges. Since an ion thermalizes and forms a snowball in a short distance from the ionization site,~\cite{Callicoatt-1996} the probability of finding the electron bubble at a distance between $r$ and $r+dr$ from the ion snowball can be written as
\begin{equation}\label{bubble-ion distance}
\frac{32}{\pi^{2}\xi^{3}}\texttt{exp}(-4r^{2}/\pi\xi^{2})r^{2}dr,
\end{equation}
where $\xi$ denotes the mean electron-ion separation distance. The chance that an ionized electron can escape from recombination with its parent ion under an applied field strongly depends on $\xi$. To this end, $\xi$ can be regarded as a measure of how easily one may extract the ionized electrons from the track of the primary energetic electron. This $\xi$ is of practical importance and is the quantity to be determined in the following sections.

\begin{figure}[htb]
\includegraphics[scale=0.42]{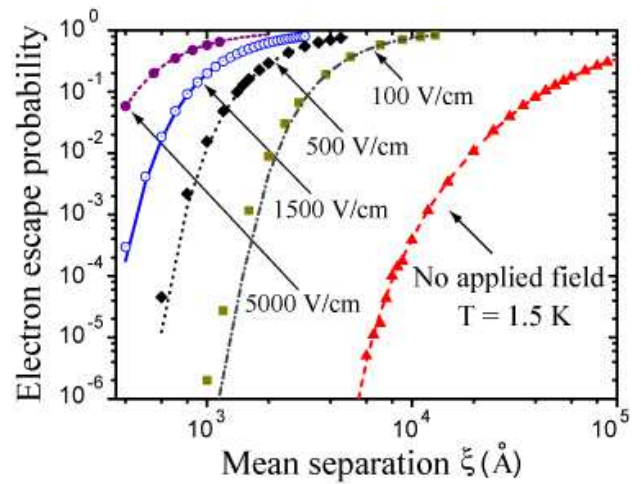}
\caption{(color online). Probability of an electron escaping recombination with its parent ion as a function of the mean electron-ion separation $\xi$ at T=1.5~K. The red triangles represent the Monte Carlo simulation with zero external electric field. The red dashed curve is calculated based on Eq.~(\ref{Onsager}) for zero electric field. The grey squares, the black diamonds, the blue open-circles and the purple solid circles represent the Monte Carlo simulation with 100~V/cm, 500~V/cm, 1500~V/cm, and 5000~V/cm external electric field, respectively. The escape probability has no appreciable temperature dependence when the external field is strong. The grey dash-dotted curve, the black dotted curve, the blue solid curve, and the purple short-dashed curve are calculated based on Eq.~(\ref{e-escape-boundary}) at the same fields with the corresponding Monte Carlo simulation.}\label{Escape probability}
\end{figure}

Let us first consider the case with no externally applied field. Above 1~K the thermalized electron bubble and the ion snowball, when their separation is not too small, both undergo a diffusive motion in the liquid due to the collisions with the quasi-particles. Meanwhile, the Coulomb attraction between the electron and the ion will cause them to move toward each other. To calculate the chance that the electron bubble diffuses away (escapes) from recombining with the positive ion, a 3D Monte Carlo simulation is performed. In this simulation, we consider the motions of an electron bubble and a He$^{+}$ ion that are initially placed at $\vec{r}$ and at the origin, respectively. In a time step ${\Delta}t$, the electron and the positive ion first move in random directions by a distance $\sqrt{6D_{el}{\Delta}t}$ and $\sqrt{6D_{ion}{\Delta}t}$. Here $D_{el}$ and $D_{ion}$ are the diffusion coefficients of the electron bubble and the He$^{+}$ ion in liquid helium. The diffusion coefficient $D$ is related to the mobility $\mu$ by the Einstein relation $D={\mu}k_{b}T/e$ and thus can be calculated using the tabulated mobility data.~\cite{Donnelly} To account for Coulomb attraction, the electron bubble and the ion snowball also move toward each other by distances $v_{el}{\Delta}t$ and $v_{ion}{\Delta}t$, respectively, following every step of their random movements. The particle velocity $v_{el(ion)}$ depends on the electron-ion distance $|\vec{r}_{el}-\vec{r}_{ion}|$
\begin{equation}\label{bubble-ion velocity}
v_{el(ion)}=\mu_{el(ion)}\cdot\frac{1}{4\pi\epsilon}\frac{e}{|\vec{r}_{el}-\vec{r}_{ion}|}.
\end{equation}
In the simulation, the inertial effect is ignored since the particle motions are over-damped in the temperature region (above 1~K) considered here when the driving field is not too strong.~\cite{WGuo-bubble} When the separation between the electron and the ion is larger than $10^{6}$~{\AA}, the electron is considered to have escaped from the ion since at this distance the Coulomb potential energy is much smaller than $k_{B}T$;
\begin{figure}[htb]
\includegraphics[scale=0.41]{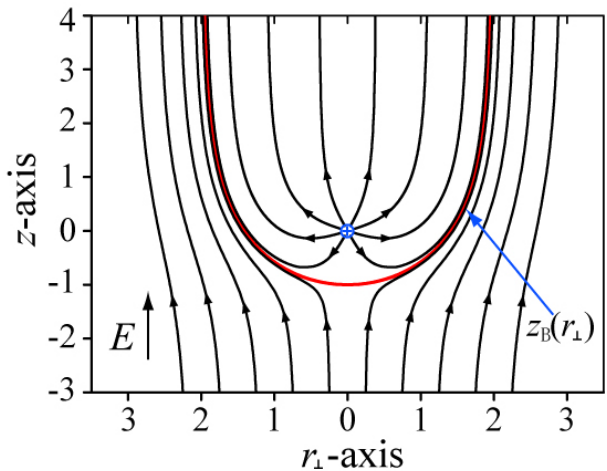}
\caption{(color online). The electric field lines near the He$^{+}$ ion in an applied field $E$. The ion is placed at the origin and the field $E$ is along the $z$-axis. The units of the $z$-axis and the cylindrical radial $r_{\bot}$-axis are both $\sqrt{e/4\pi{\epsilon}E}$. The boundary surface $z=z_{B}(r_{\perp})$ (red curve) separates the ion's field lines from those belong to the applied field $E$.}\label{Field lines}
\end{figure}
whereas if the separation is less than $100$~{\AA}, the charge pair is considered to have recombined. We then consider an ensemble of $10^{6}$ such bubble-ion pairs with starting location $r$ distributed according to Eq.~(\ref{bubble-ion distance}), and find the probability that a bubble can escape the positive ion. The calculated electron escape probability as a function of the mean range $\xi$ at $T=1.5$~K is shown as red triangles in Fig.~\ref{Escape probability}. In fact, since the Einstein relation connects diffusion and mobility, the electron escape probability is a function of the single parameter ${\xi}k_{B}T/e^{2}$. For the case that the initial separation of the positive and negative ion is $r$, the electron escape probability, as first derived by Onsager,~\cite{Onsager-paper} is given by $\texttt{exp}(-e^{2}/4\pi{\epsilon}rk_{B}T)$. As a result, the average escape probability is
\begin{equation}\label{Onsager}
P_{escape}=\frac{32}{\pi^{2}\xi^{3}}\int^{\infty}_{0}\!\texttt{exp}(\frac{-4r^{2}}{\pi\xi^{2}})\texttt{exp}(\frac{-e^{2}}{4\pi{\epsilon}rk_{B}T})r^{2}\texttt{d}r.
\end{equation}
The red dashed curve in Fig.~\ref{Escape probability} shows the calculated escape probability based on Eq.~(\ref{Onsager}). Our Monte Carlo simulation result agrees well with one obtained using Eq.~(\ref{Onsager}), which confirms the validity and accuracy of our Monte Carlo method. As can be seen, if the electron-ion mean separation $\xi$ is not much greater than 10$^{4}$~{\AA}, the electrons and ions predominantly recombine in the absence of externally applied fields.

When an external electric field $E$ is applied (assumed to be along the $z$-axis), along with the random walk and the motion towards each other, in each time step the electron and the ion also move along the $z$-axis by step-lengths $-\mu_{el}E\cdot{\Delta}t$ and $\mu_{ion}E\cdot{\Delta}t$, respectively. We adapted our Monte Carlo method to determine the electron escape probability since Eq.~(\ref{Onsager}) does not apply in this case. The obtained escape probability as a function of the electron-ion mean separation $\xi$ at $T=1.5$~K under fields 100~V/cm, 500~V/cm, 1500~V/cm, and 5000~V/cm are shown in Fig.~\ref{Escape probability} as grey squares, black diamonds, blue open-circles, and purple solid-circles, respectively. As one can see, the electron escape probability is greatly enhanced. Another observation in the simulation is that, for a field that is not too weak, the calculated escape probability versus the mean separation $\xi$ does not depend much on temperature. This can be understood in the following way. The electric field lines near the He$^{+}$ ion in the applied field $E$ are calculated and shown in Fig.~\ref{Field lines}. The ion is placed at the origin and the length scale in the figure is in units of $\sqrt{e/4\pi{\epsilon}E}$. There is a boundary surface $z=z_{B}(r_{\perp})$ which separates the field lines that are created by the ion from those belonging to the applied field. In the moving frame of the ion, the electron moves along the field lines if its motion is ideally over-damped. An electron that is initially placed inside the volume enclosed by the boundary surface $z=z_{B}(r_{\perp})$ will trace the field line back and recombine with the ion; otherwise the electron escapes if it is placed outside the boundary surface. The real motion of the electron at the temperature considered here is, however, a zigzagged path that may shift from one field line to nearby field lines. Moving along the field line by a distance $l$, the electron may shift away from the field line by a distance on the order of $\sqrt{6D(l/{\mu}E)}$. When the field is reasonably strong, the zigzag motion can only affect the destiny of an electron that is initially located very close to the boundary surface. As a result, the temperature effect on the electron escape probability is weak. Our inference is confirmed by Ghosh's experimental observation~\cite{Ambarish-thesis} that the field-dependence of the electron current extracted from a beta source immersed in liquid helium does not depend much on temperature when the field is higher than a few tens of Volts per centimeter. Based on the above reasoning, a much simpler and faster estimation of the electron escape probability can be made by integrating the probability that the electron initially appears outside the boundary surface
\begin{align}\label{e-escape-boundary}
P_{escape}&=1-\int^{\infty}_{-1}\!\texttt{d}z_{B}\int^{r_{\perp}(z_{B})}_{0}\!2{\pi}r_{\perp}\texttt{d}r_{\perp}\cdot\frac{8}{\pi^{3}{\xi'}^{3}}\nonumber\\
          &\cdot\texttt{exp}\left\{-\frac{4(z_{B}^{2}+r_{\perp}^{2})}{{\pi}{\xi'}^{2}}\right\}.
\end{align}
Here $\xi'$ is the electron-ion mean separation $\xi$ in unit of $\sqrt{e/4\pi{\epsilon}E}$. The grey dash-dotted curve, the black dotted curve, the blue solid curve, and the purple short-dashed curve in Fig.~\ref{Escape probability} show the calculated electron escape probability based on Eq.~(\ref{e-escape-boundary}) under fields 100~V/cm, 500~V/cm, 1500~V/cm, and 5000~V/cm, respectively. The difference between the above simple calculation and the Monte Carlo simulation becomes appreciable when the field is below about 50~V/cm at 1.5~K.

\begin{figure}[htb]
\includegraphics[scale=0.42]{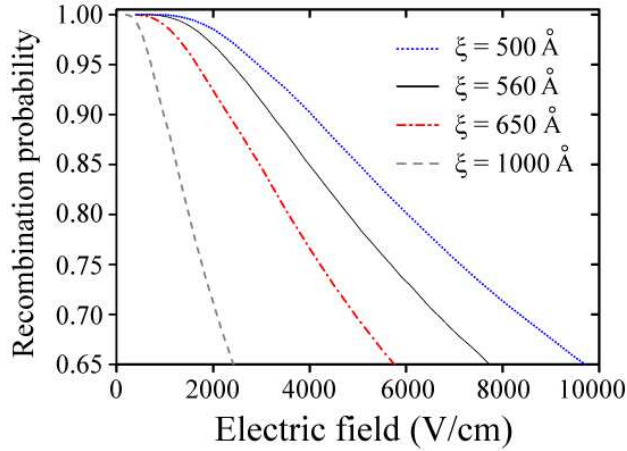}
\caption{(color online). Monte Carlo simulation of the electron-ion recombination probability as a function of the applied electric field at 1.5~K. The blue dotted line, the black solid line, the red dash-dotted line and the grey dashed line are for electron-ion mean separation of 500~{\AA}, 560~{\AA}, 650~{\AA} and 1000~{\AA},
respectively.} \label{Quenching factor}
\end{figure}

We can also calculate the field dependence of the electron escape probability at an assumed mean separation $\xi$. In Fig.~\ref{Quenching factor}, instead of showing the electron escape probability, we plot the probability that the electron recombines with the He$^{+}$ ion as a function of the applied electric field at four assumed mean separation distances (500~{\AA}, 560~{\AA}, 650~{\AA} and 1000~{\AA}). The recombination probability decreases with increasing field and its field dependence depends sensitively on the value of the electron-ion mean separation $\xi$. As discussed in the previous section, the number of prompt scintillation photons emitted along the track of an energetic electron in liquid helium is proportional to the number of electron-ion pairs that recombine after ionization. The ratio $q$ of the prompt scintillation light with applied field $E$ to the prompt light without field is thus a measure of the electron-ion recombination probability under field $E$. As a result, measuring $q$ at different applied fields and comparing the result with the simulation discussed above allows an accurate determination of the electron-ion mean separation $\xi$.

\section{Experiment and results}
\begin{figure}[htb]
\includegraphics[scale=0.42]{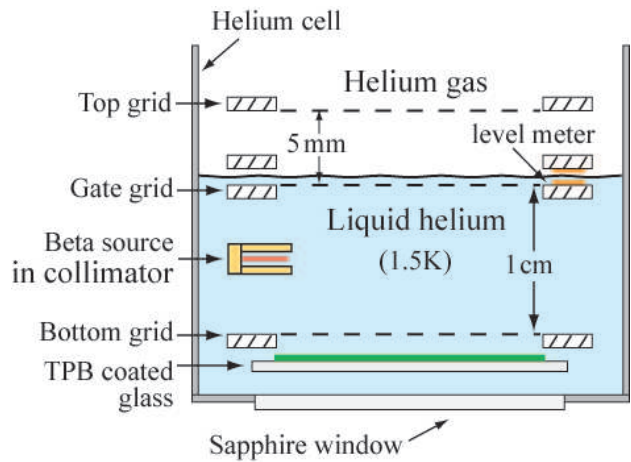}
\caption{(color online). Schematic diagram showing the experimental setup.} \label{Setup}
\end{figure}

A schematic diagram of the experimental setup is shown in Fig.~\ref{Setup}. A helium cell is thermally linked to a pumped liquid helium reservoir and its temperature was controlled to be $1.5\pm0.02$~K. The device held inside the helium cell is composed of four Cirlex rings~\cite{Cirlex}. Three of the rings are covered with nickel mesh grids (1~inch in diameter, transparency 90\%) and are used as cathode (bottom grid), gate (gate grid), and anode (top grid). The distance from the bottom grid to the gate grid and from the gate grid to the top grid are 1~cm and 0.5~cm, respectively. Voltages in both polarities with amplitude up to 1500~V can be applied to each of these grids. The fourth ring is placed at about 0.5~mm above the gate grid by nylon washers. Three separated copper patches on the fourth ring form three capacitors with the corresponding patches on the gate ring. These capacitors serve as level meter such that we can condense helium into the cell with liquid level just above the gate grid. A $^{90}$Sr needle beta source in a copper tube collimator with 0.2~mm inner diameter is held at the middle between the bottom and the gate grids to produce energetic electrons in liquid helium. The beta emission event rate is about 20 per second at the exit of the collimator, measured using a Geiger counter. The emitted energetic electrons move approximately horizontally in the cell, producing ionization and excitation along their tracks. The low event rate allows us to study the scintillations from individual beta tracks on a event-by-event basis. In order to convert the EUV scintillation photons to the visible range for convenient detection, a 1~mm thick glass disc was placed beneath the bottom grid with its upper face coated by a layer of tetraphenyl butadiene (TPB) organic fluor. TPB has been previously shown to be a very efficient fluor for converting helium scintillation into blue light.~\cite{Dan-TPB}

This experiment was designed with multiple goals. First, by varying the voltages applied to the bottom and the gate grids, the field dependence of the prompt scintillations produced along the tracks of the beta particles can be studied. This will allow us to determine the electron-ion separation distance. Second, the ionized electrons extracted from the track of a beta particle can be drifted into the helium gas phase to produce proportional scintillation when a high positive voltage is applied to the top grid. A brief discussion on detecting the extracted electrons will be given in the next section.

\begin{figure}[htb]
\includegraphics[scale=0.42]{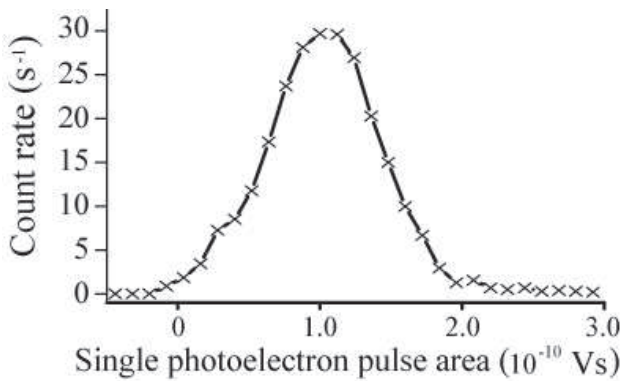}
\caption{PMT dark-count pulse area histogram. Data were collected in $1.2\times10^{-11}$~Vs bins. The oscilloscope trigger level was set to 4~mV.} \label{dark-count}
\end{figure}

To detect the converted blue photons, a 5-cm diameter photomultiplier tube (PMT)~\cite{PMT} was placed 8~cm below the helium cell on a base and voltage divider mounted at room temperature inside the vacuum shield of the cryostat. The fraction of the EUV light that reached the TPB coated glass was roughly 15\%. About 2.4\% of the converted blue light was collected by the PMT. The overall light collection efficiency is thus about 0.4\%. The signal from the PMT was read out by a Tektronix TDS~5104 digital oscilloscope. Thermionically emitted individual electrons from the PMT photocathode (dark counts) provided us a means to characterize the PMT response to single photoelectrons. We analyzed the dark counts with the PMT alone inside the cryostat with 1300~V applied to PMT photocathode. The pulse-area histogram of the dark counts is shown in Fig.~\ref{dark-count}. The particular area bin ($1.04\times10^{-10}$~Vs) that corresponds to the peak in the histogram plot is defined as the characteristic pulse area of the PMT response to a single photon.

\begin{figure}[htb]
\includegraphics[scale=0.42]{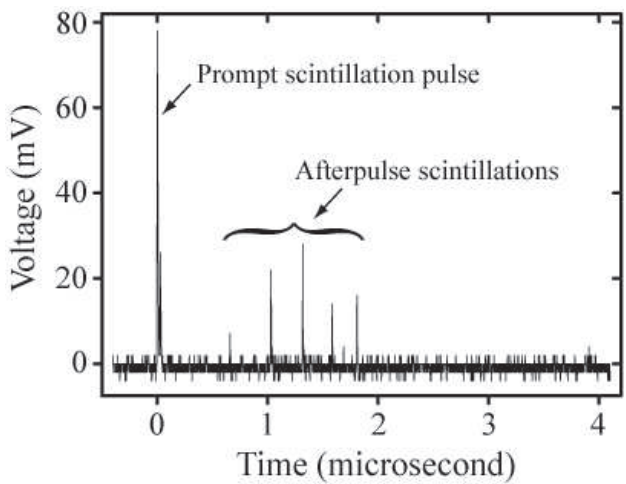}
\caption{Typical trace for a beta emission event in liquid helium. The oscilloscope trigger level was set to 50~mV.} \label{typical trace}
\end{figure}

Each energetic electron produced by the beta source moves through the liquid and produces a prompt scintillation pulse and some afterpulse scintillation. The prompt scintillation pulse consists of many photons, whereas the afterpulse scintillations should essentially be individual single photons spread over a long time. As a result, we may trigger on the prompt pulse to distinguish a beta emission event from background dark counts by setting the oscilloscope trigger level higher than the average pulse height for single photons. In Fig.~\ref{typical trace}, a typical trace for a beta emission event is shown with 50~mV trigger level. The pulse areas of individual afterpulses are analyzed. The generated area histogram shows a similar profile as the one for the dark counts (Fig.~\ref{dark-count}), which confirms that the afterpulse scintillations are indeed individual single photons. The number of photons in the prompt pulse is proportional to the initial energy of the particle. A histogram of the areas of the prompt scintillation pulses thus reflects the energy spectrum of the emitted beta particles.

\begin{figure}[htb]
\includegraphics[scale=0.4]{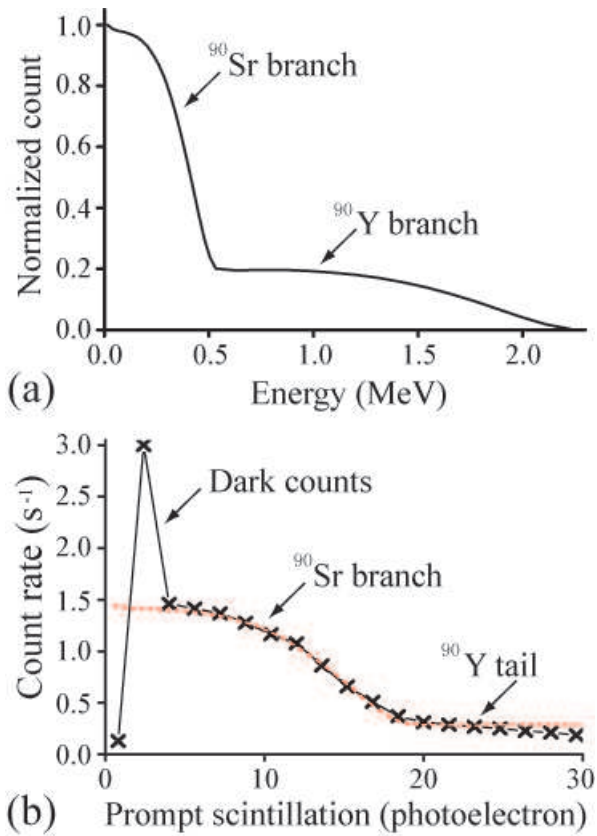}
\caption{(color online). (a)~Theoretical energy spectra of beta particles emitted from a $^{90}$Sr/$^{90}$Y source. (b)~Histogram plot of prompt scintillation pulse area in terms of photoelectrons. Data were collected in $1.6\times10^{-10}$~Vs (e.g. 1.54 photoelectrons) bins. The oscilloscope trigger level was set to 35~mV. The dashed red curve is scaled from the theory curve shown in (a).} \label{spectrum}
\end{figure}

A $^{90}$Sr atom undergoes $\beta^{-}$ decay to an electron, an anti-neutrino and the yttrium isotope $^{90}$Y with decay energy of 0.546 MeV. The half life of $^{90}$Sr is 28.8 years. $^{90}$Y also undergoes $\beta^{-}$ decay with half life of 64 hours and decay energy 2.28 MeV distributed among an electron, an anti-neutrino, and $^{90}$Zr (zirconium), which is stable. The theoretical energy spectra of the beta particles emitted from a $^{90}$Sr/$^{90}$Y source, drawn from the Radiation Dose Assessment Resource (RADAR),~\cite{Sr-90} is shown in Fig.~\ref{spectrum}~(a). The endpoint energy of the $^{90}$Y branch in the energy spectrum is 2.28~MeV. A beta particle with this maximum energy can traverse in liquid helium a distance of about 4.5~cm. Such beta particles will collide on the cell walls or the support rods of the grids due to the limited size of the cell. In the experiment, the beta particles from the $^{90}$Sr branch, with track length $\lesssim1$~cm, are our object of study. Over 50~thousand prompt scintillation pulses are analyzed. By dividing the prompt scintillation pulse area by the PMT characteristic single photon pulse area, the number of photoelectrons detected in each prompt pulse is calculated. The obtained prompt scintillation pulse area histogram, in terms of single photoelectrons, is shown in Fig.~\ref{spectrum}~(b). This histogram plot was made with trigger level 35~mV. Any pulses with amplitude lower than 35~mV are not acquired by the oscilloscope, which leads to a sharp lower cutoff in the histogram. The contribution to the count rate from the $^{90}$Sr branch is clearly seen with an endpoint roughly around 18~photoelectrons. The tail part in the histogram at large photoelectron number comes from the $^{90}$Y decay. The rise of the count rate in Fig.~\ref{spectrum}~(b) at small photoelectron number is due to the counting of PMT dark counts with amplitude higher than 35~mV. Since the dark count rate falls exponentially at pulse area larger than one characteristic single photon pulse area, the major part of the $^{90}$Sr branch in the histogram is not affected by the dark counts. The red curve in Fig.~\ref{spectrum}~(b) is scaled from the theoretical energy spectrum curve shown in Fig.~\ref{spectrum}~(a). A good agreement between the theoretical and measured spectra is seen.

When voltages are applied to the bottom and the gate grids to create an electric field $E$ in liquid helium, some ionized electrons along the track of an emitted beta particle are extracted. As a result, the number of photons in the prompt scintillation pulse decreases, which in turn leads to a change in the pulse area histogram plot. We may think about this histogram profile deformation in the following way. First, those events that at zero applied field lead to the production of prompt scintillations comprising $N$ photons now only produce $qN$ prompt photons under the applied field $E$. Here $q$ refers to the field quenching ratio ($q\lesssim1$). As a consequence, for any points on the prompt-pulse area histogram curve at zero field, their horizontal coordinates need to be multiplied by $q$ when the field $E$ is applied. In some sense, the histogram profile ``shrinks'' by a factor of $q$ along the horizontal axis. On the other hand, the sum of events per unit time is indeed the radioactivity of the source which is statistically a constant that does not depend on externally applied field. Hence at the same time the vertical coordinates of the points on the histogram curve need to be multiplied by $1/q$. In short, converting a prompt scintillation pulse area histogram at zero field to one at finite field $E$ and acquired over an equal time interval with same radioactive source, one just needs to contract the histogram profile by $q$ along the horizontal axis and then stretch the profile by $1/q$ along the vertical axis. Such a deformation of the histogram profile is clearly seen experimentally.

\begin{figure}[htb]
\includegraphics[scale=0.4]{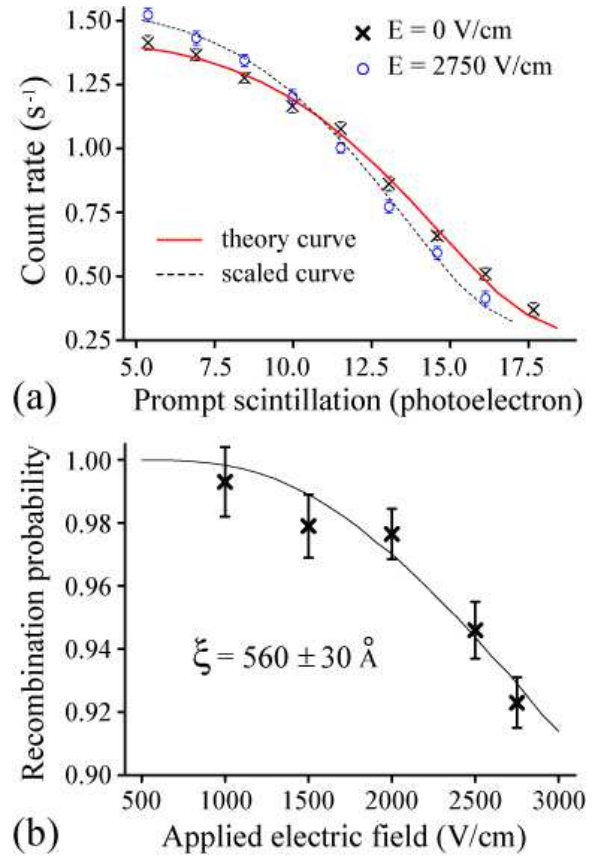}
\caption{(color online). (a)~Histogram plot of prompt scintillation pulse area in terms of photoelectrons. The black crosses and the blue circles represent data obtained at zero applied field and at $E=2750$ V/cm, respectively. The red solid curve is scaled from the theory spectrum (Fig.~\ref{spectrum}~(a)) to get the best fit to the zero-field data. The black dashed curve is scaled from the solid curve to fit the data at $E=2750$ V/cm. All the data were collected in $1.6\times10^{-10}$~Vs (e.g. 1.54 photoelectrons) bins with oscilloscope trigger level set to 35~mV. (b)~Electron-ion recombination probability as a function of the applied electric field. The crosses represent the experimental data. The solid line is the simulated curve that gives the best fit to the experimental data. The obtained electron-ion mean separation distance $\xi=560\pm30$~{\AA}.} \label{spectrum-field}
\end{figure}

In Fig.~\ref{spectrum-field}~(a), we show the data obtained for the prompt scintillation pulse area histogram at zero applied field and at $E=2750$~V/cm as black crosses and blue circles, respectively. Only the $^{90}$Sr portion of the histogram is plotted for a clear view of the profile deformation. The red solid curve is scaled from the theory spectrum (Fig.~\ref{spectrum}~(a)) in a way to get the best fit to the zero-field data. The difference between the zero-field histogram profile and the one with 2750~V/cm field is obvious. The field quenching ratio $q$ can be determined as follows. For a given $q$ value, we contract the red curve by $q$ along the horizontal axis and then stretch it by $1/q$ along the vertical axis. The difference between the obtained curve (shown in Fig.~\ref{spectrum-field}~(a) as the black dashed curve) and the data at field $E=2750$~V/cm is calculated by summing up the square of the vertical distance from each data point to the obtained curve. A minimum-$\chi^{2}$ fit~\cite{Chi-2} gives $q=(92.4\pm0.8)\%$ for field $E=2750$~V/cm. Note that in determining the quenching ratio $q$, only the portion of the data shown in Fig.~\ref{spectrum-field}~(a), above 5 photoelectrons, is used, in order that the PMT dark counts not affect the result. Similar analyzes were performed for different applied electric fields between the bottom and the gate grids.

As was discussed previously, the field quenching ratio $q$ is equal to the probability that the ionized electrons recombine with their parent ions under the applied field. In Fig.~\ref{spectrum-field}~(b), the obtained electron-ion recombination probability data as a function of the applied electric field are shown as black crosses. To compare the data with the result of the Monte Carlo simulation described in previous section (see Fig.~\ref{Quenching factor}), we performed a standard $\chi^{2}$-analysis. For a given electron-ion mean separation distance $\xi$, the difference between the data and the simulated curve is calculated by summing up the square of the vertical distance from each data point to the simulated curve weighted by the square variation of the data point. We then vary $\xi$ in 10~{\AA} steps to minimize this difference. The solid line in Fig.~\ref{spectrum-field}~(b) is the simulated curve that gives the best fit to the experimental data. The obtained electron-ion mean separation distance $\xi$ is $560\pm30$~{\AA}. The reduced-$\chi^{2}$ of this fit is 0.85. Note that since the ionized electrons thermalize through collisions with helium atoms instead of with quasi-particles in superfluid helium, the electron-ion mean separation we obtained at 1.5~K should remain unaltered even at very low temperatures. It is worth mentioning that in a recent experiment conducted by Ito \emph{et al.}~\cite{Ito}, similar electron-ion separation (track width parameter $b\sim600$~{\AA}) was found along the tracks of $\alpha$ particles stopped in superfluid helium.

\section{Discussion}
The fraction of the ionized electrons extracted in the current experiment is limited by the voltages that can be applied to the cables. A modified design of the electric feedthroughs on our helium cell will allow us to achieve a drift field over 5000~V/cm in the liquid. Under such a field, about 23\% of the ionized electrons can be extracted according to our Monte Carlo simulation. Even higher extraction efficiency may be expected at temperatures lower than 1~K since the ionized electrons may stray away from the curved field lines near the ions due to inertial effect and escape the recombination. An electron bubble approaching the helium free surface induces a negative image charge on the surface and can become trapped below the interface due to the combined effects of this image charge and the driving electric field.~\cite{Ancilotto-1994} The trapping life time $\tau$ depends on temperature and the driving field. At 1.5~K and under a driving field of a few thousand volts per centimeter, $\tau$ is less than a microsecond.~\cite{Schoepe-1973} The extracted electrons can thus be quickly pulled across the interface into the helium gas.

To read out the electron signal of a two-phase liquid/gas helium detector, Gas Electron Multipliers (GEMs) or Thick Gas Electron Multipliers (THGEMs) can be used.~\cite{Buzulutskov-2005,Buzulutskov-2007, Breskin} Detecting the extracted electrons using GEMs has already been studied experimentally by the ``e-bubble'' collaboration.~\cite{Ju-2007,Ju-2007-Chin} A disadvantage of the GEM is that since it relies on a breakdown for electron gain, it gives poorer energy resolution than proportional scintillation. However, the very slow electron drift speed, normally a disadvantage for liquid helium, can actually be an advantage allowing better energy resolution. Because of the slow electron drift speed, the electrons will arrive at the GEM one at a time, easily distinguishable due to the excellent GEM timing resolution. We may operate the GEM in a single electron detection mode, counting single electron pulses instead of using the pulse sizes to determine the event energy. At the same time the GEM (or stack of GEMs) could also be used to read out the scintillation signal. The side of the GEM facing the liquid could be coated with Cesium Iodide (CsI)~\cite{Breshin-CsI} or other photocathode material so as to be sensitive to the prompt scintillation light.

Alternatively one may detect the extracted electrons by driving them in the gas phase with strong electric field to produce proportional scintillation light as is commonly done in Argon, Krypton, and Xenon two-phase detectors.~\cite{Conde-1967,Aprile-2004,Benetti,Bolozdynya} Electrons moving in the helium gas collide with helium atoms. Under an appropriate field, an electron may gain a kinetic energy exceeding the excitation threshold of helium atoms between collisions. Excited helium atoms can thus be produced along the path of each individual electron due to electron impact. The excited helium atoms in spin-singlet states can react with surrounding helium atoms and decay to the ground state by emitting an EUV photon. The number of EUV photons produced along the electron path is proportional to the path length of the electron in helium gas. As a consequence, following the prompt scintillation pulse of each beta emission event, we expect to see a burst of proportional scintillation due to the electrons being pulled through the helium gas. Electroluminescence might also be produced using very high ($\sim$1-10~MV/cm) fields in liquid helium as has been observed in liquid Ar~\cite{Lightfoot} and liquid Xe~\cite{XENON100}. This could allow electrons to be individually detected, while not subjecting gaseous helium to an electric field. The high field might be produced with thin wires or a GEM. A detailed study on detecting the proportional scintillation will be reported in a future paper. 

With efficient detection of electrons, achieved through gas-phase or liquid-phase amplification, a very low energy threshold ($\sim${keV}) may plausibly be achieved. A two-phase helium detector might be used for low mass (0.1 to 10 GeV) WIMP detection, with the advantage that helium nuclei are kinematically well-matched to the WIMPs in this mass range and, compared to a target of heavier atoms with the same energy threshold, would be sensitive to a larger fraction of the dark matter velocity distribution. Metastable triplet helium molecules might be a third detection channel for dark matter, as discussed by McKinsey \emph{et al}.~\cite{McKinsey-2005} and may be drifted through superfluid helium using a heat flux.~\cite{Guo-2009,Guo-2010} In order to better determine the applicability of liquid helium for a low-mass WIMP search, further research is needed to determine the scintillation, ionization and triplet molecule yields for nuclear recoils in liquid helium. The ratio of ionization to scintillation, used to good effect in liquid Xenon~\cite{Gaitskell-2004,Aprile-2010, XENON10,ZEPLIN-II,ZEPLIN-III} and liquid Argon~\cite{Lippincott-2008,Brunetti-2005} WIMP search experiments for discriminating between electron and nuclear recoils, may also allow discrimination in liquid helium. The ratio of singlet to triplet molecules may also allow discrimination, as has been shown in liquid Argon~\cite{Lippincott-2008, Benetti} and liquid Xenon~\cite{Kwong, Yamashita}. Because of the low mass of the helium nucleus, the Lindhard effect~\cite{Lindhard} is relatively mild, and scintillation and ionization yields are only expected to be modestly reduced for nuclear recoils in comparison to electron recoils.~\cite{Mayet} The possibility that liquid helium might be used for dark matter detection, using scintillation, charge, and metastable He$^{*}_{2}$ excimer signals, will be discussed further in a future publication.

\section{Summary}
The prompt scintillation produced by individual beta emission events from a $^{90}$Sr source in liquid helium has been investigated. The field dependence of the prompt scintillation light can be used to determine the mean electron-ion separation distance for charge pairs produced along the beta tracks. By comparing the observed field-induced quenching of the prompt light with our Monte Carlo simulations, a mean electron-ion separation of about $560\pm30$~{\AA} is obtained.

\begin{acknowledgements}
The authors would like to thank Prof. H. J. Maris and Prof. G. M. Seidel for helpful discussions. This work was supported by Yale University.
\end{acknowledgements}

\end{document}